# A Scenario-based Stochastic Model of using BESS-based Virtual Transmission Lines in Day-Ahead Unit Commitment

Qiushi Wang
Department of Electrical and Computer Engineering
University of Houston
Houston, Texas, US

Xingpeng Li
Department of Electrical and Computer Engineering
University of Houston
Houston, Texas, US

*Abstract*—The rapid increase in renewable energy sources (RES) implementation in the power system creates more severe network congestion, which may reduce grid operation efficiency and cause renewable curtailment. Deterministic optimization for the unit commitment shows that battery energy storage system (BESS)-based Virtual Transmission Line (VTL), as an alternative to physical transmission lines, can offer a quick solution for congestion relief, reduced operational costs, and lowered renewable curtailment. This paper aims to evaluate the benefits of VTL when considering Renewable Energy Sources uncertainty. Particularly, this work proposes a scenario-based stochastic security-constrained unit commitment model considering VTL, referred to as SSCUC-VTL. It incorporates the forecast error of RES into the commitment decision for systems with VTL. The performance of applying the VTL strategy is compared to that of adding a new physical transmission line and a standalone BESS. A case study has been conducted on an enhanced IEEE 24-bus test system. The simulation results demonstrate that VTL provides 23% more operational cost reduction than the physical transmission line, and up to 67% more congestion relief than the standalone BESS in a power system with solar and wind generation.

## NOMENCLATURE

| | |
|---|---|
| $g$ | Transmission element (line or transformer) index. |
| $G(n)$ | Set of generators at bus n. |
| $n$ | Bus index. |
| $N$ | A set of all buses. |
| $k$ | Transmission element (line or transformer) index. |
| $K$ | Set of all transmission elements. |
| $K(n-)$ | Set of branches with bus n as the to-bus. |
| $K(n+)$ | Set of branches with bus n as the from-bus. |
| $r$ | Solar generation index. |
| $R(n)$ | Set of all solar generators at bus n. |
| $w$ | Wind generation index. |
| $W(n)$ | Set of all wind generators at bus n. |
| $e$ | Battery index. |
| $ES(n)$ | Set of all battery storage systems at bus n. |
| $ES(vt)$ | A set of all battery-based virtual transmission systems. |
| $vt$ | Virtual transmission line index. |
| $VT$ | A set of all Virtual transmission lines. |
| $t$ | Period index. |
| $T$ | A set of all Periods. |
| $s$ | Scenario index. |
| $S$ | A set of all scenarios. |
| $\pi_s$ | Probability of scenarios. |
| $x_k$ | The reactance of the transmission element. |
| $C_g$ | Linear cost for generator g. |
| $C_g^{NL}$ | No-load cost for generator g. |
| $C_g^{SU}$ | The start-up cost for generator g. |
| $C_{res}$ | The penalty cost for solar and wind curtailment |
| $d_{nts}$ | Predicted load demand of bus n in the period t for scenario s. |
| $BigM$ | A big real number. |
| $P_g^{min}$ | The minimum capacity of generator g. |
| $P_g^{max}$ | Maximum capacity of generator g. |
| $R_g^{hr}$ | Generator ramping rate limit |
| $P_k^{max}$ | Emergency thermal line limit for line k. |
| $u_{gt}$ | Commitment status of unit g in the period t. |
| $v_{gt}$ | Start-up variable of generator g in the period t. |
| $\theta_{kts}$ | Phase angle difference between the from-end and to-end of line k in the period t for scenario s. |
| $P_{gts}$ | The output of generator g in the period t for scenario s. |
| $P_{kts}$ | Flow in line k in the period t for scenario s. |
| $P_{rts}$ | The output of solar generators in the period t for scenario s. |
| $P_{wts}$ | The output of wind generators in the period t for scenario s. |
| $P_{rts}^C$ | The curtailed solar generation in the period t for scenario s. |
| $P_{wts}^C$ | The curtailed wind generation in the period t for scenario s. |
| $P_{ets}^c$ | Charging rate of battery e in the period t for scenario s. |
| $P_{ets}^d$ | Discharging rate of battery e in the period t for scenario s. |
| $E_{ets}$ | Energy storage energy level in the period t for scenario s. |
| $u_{ets}^c$ | 1 indicates charging mode; otherwise, 0. |
| $u_{ets}^d$ | 1 indicates discharging mode; otherwise, 0. |
| $E_e^{min}$ | Minimum energy storage energy level. |
| $E_e^{max}$ | Maximum energy storage energy level. |
| $P_e^{c,max}$ | Maximum energy storage charge rate. |
| $P_e^{d,max}$ | Maximum energy storage discharge rate. |
| $\eta_e^c$ | Charging efficiency. |
| $\eta_d^d$ | Discharging efficiency. |
| $\rho_r$ | The solar penalty coefficient |
| $\rho_w$ | The wind penalty coefficient |
| $\Delta T$ | Length of a time interval. |
| $LMP_{nt}$ | Locational marginal price (LMP) at bus n in the period t. |

## I. INTRODUCTION

The virtual transmission line (VTL), also known as the virtual power line (VPL), is an innovative operational strategy for energy storage systems (ESSs) that alleviates network congestion and promotes the use of renewable energy resources (RES). A VTL includes at least one pair of ESSs at two different network nodes: for example, one on the supply side and the other on the load side. The idea is to cooperate the BESSs on two sides of the transmission line, so that 1) the BESS on the source end charges when the line is congested and discharges when there is available capacity; 2) the BESS on the load end discharges when the line is congested and charges when there is available capacity. Reference [1] summarized that there are around 3 GW of VTL initiation globally in 2020, and highlighted the ongoing projects in France, Italy, Australia, and Germany, which show that the VTL helps reduce electricity prices, defer network upgrades, and improve system reliability.

Battery energy storage systems (BESS) have been selected for VTL applications in most of the literature. Reference [2] presented a mixed-integer linear programming (MILP) unit commitment (UC) model for the BESS-based VTL to optimize the available renewable power and relieve system congestion. Reference [3] proposed a security-constrained UC (SCUC) model for a BESS-based mobile VTL to increase transmission capacity and reduce total operation cost. Reference [4] conducted a comprehensive evaluation of the performance and economic benefit of BESS-based VTL against physical transmission line, standalone BESS, and transmission line switching strategy. The study concluded that the BESS-based VTL helps achieve a balanced solution in terms of total grid operation cost, load payment, congestion relief, computational time, and renewable curtailment reduction.

The basic approach for the existing VTL studies is the deterministic SCUC, which assumes fixed load and generation levels for the optimization problem. A limitation of the deterministic approach is that it fails to account for the uncertainty and variability of wind and solar generation. Stochastic optimization needs to be introduced to model the VTL-related SCUC problem for BESS-based VTL, as BESS-based VTL has been proposed as an alternative to a physical transmission line for systems with high renewable penetration. In such systems, the commission and dispatch decisions of the BESS-based VTL depend largely on the fluctuations of wind and solar generation.

References [5, 6] provided an overview of fundamentals and recent developments in the scenario-based stochastic UC problem. The authors summarized that most of the research falls into four programming methods: basic two-stage stochastic programming, basic multi-stage stochastic programming, security-constrained robust optimization, and chance-constrained formulation. The modeling of uncertainty could be categorized as scenarios, uncertainty sets, and probabilistic constraints. Two common combinations of uncertainty modeling and stochastic optimization methods are scenario-based stochastic programming and uncertainty-set-based robust optimization. Reference [7] presented two modeling approaches for reducing the computational burden of stochastic UC formulation. One is to relax the integrality constraints for fast-start units in a two-stage stochastic unit commitment formulation. This approach enables the commitment decision for fast-start generators to be made simultaneously with dispatch decisions for slow-start generators in the second stage.

Since scenario-based stochastic programming can provide a reliable solution that considers RES uncertainty, it has been utilized to evaluate planning strategies and emerging applications in both long-term [8, 9] and short-term planning [10, 11, 12]. While a few papers discussed the VTL as an innovation in BESS control strategy, little literature has evaluated the performance of VTL using a Stochastic SCUC (SSCUC) model to include the influence of RES uncertainty.

In this paper, we propose a scenario-based SSCUC model with VTL, SSCUC-VTL, that incorporates the impact of uncertainty in renewable energy sources. The SSCUC-VTL model is used to evaluate the performance of VTL in comparison to various congestion mitigation strategies. We define the SSCUC model without any congestion mitigation scheme as the SSCUC-Base, the SSCUC model with a new physical transmission line as the SSCUC-PT, and the SSCUC model with a standalone BESS as the SSCUC-BESS.

The remainder of this paper is organized as follows. Section II describes the formulation of the proposed SSCUC model for VTL. Section III presents the optimization results of the IEEE 24-bus test case and compares the performance of different congestion management strategies. Finally, Section IV concludes the study and suggests future work.

## II. PROPOSED FORMULATION

The proposed SSCUC-VTL model uses mixed-integer linear programming (MILP) in a two-stage resource formulation. The uncertainty in solar and wind availability is represented by multiple scenarios S and their corresponding probabilities $\pi_s$. The decisions for traditional generators are divided from the commitment decisions for BESS-based VTL and dispatch decisions for all generations, including RES. RES curtailment is discouraged by adding a penalty in the objective function.

### A. Objective Function

The objective of the SSCUC models is to minimize the total generation cost over multiple periods, such as 24 hourly intervals for day-ahead energy scheduling, across various scenarios. The total cost of generation includes no-load cost $c_g^{SU}$, startup cost $c_g^{NL}$, weighted operation cost $c_g$, and the curtailment penalty of RES $C_{res}$. The weight of the operation cost is the probability $\pi_s$ of the associated scenario $s$. The formulation of the objective function is shown in (1)-(2).

$$minimize \sum_{g \in G} \sum_{t \in T} \left( c_g^{NL} * u_{gt} + c_g^{SU} * v_{gt} + c_g \sum_{s \in S} \pi_s P_{gts} \right) + C_{res} \quad (1)$$

where,

$$C_{res} = \rho_r \sum_{s \in S} \pi_s \sum_{r \in R} P^C_{rts} + \rho_w \sum_{s \in S} \pi_s \sum_{w \in W} P^C_{wts} \qquad (2)$$

The commitment status $u_{gt}$ and the start-up indicator $v_{gt}$ remains the same for all scenarios, as we assume all units are slow large-scale traditional generators, except for the RES and BESS units.

*B. Stochastic SCUC Constraints*

Equations (3)-(12) are the constraints for the basic SSCUC optimization model. For the generator unit commitment status $u_{gt}$ and startup indicators $v_{gt}$, definitions are given in (3)-(4), respectively. Equation (5) defines the relation between $u_{gt}$ and $v_{gt}$. Generator output limits are enforced in (6)-(7). Generator ramping rate limits are respected in (8)-(9). The line power flow equation and thermal capacity limit are presented in (9)-(10), respectively. Equation (11) ensures that the power balance is met at each node in each time interval for each scenario. The curtailment limits of solar and wind generation are defined in (12)-(13). The nodal power balance equation for the base case, with renewable generation injection and curtailment, is represented in (14).

$$u_{gt} \in \{0,1\}, \forall g, t \qquad (3)$$

$$v_{gt} \in \{0,1\}, \forall g, t \qquad (4)$$

$$v_{gt} \geq u_{gt} - u_{g,t-1}, \forall g, t \qquad (5)$$

$$P_g^{min} * u_{gt} \leq P_{gts}, \forall g, t, s \qquad (6)$$

$$P_{gts} \leq P_g^{max} * u_{gt}, \forall g, t, s \qquad (7)$$

$$P_{gts} - P_{g,t-1,s} \leq R_g^{hr}, \forall g, t, s \qquad (8)$$

$$P_{g,t-1,s} - P_{gts} \leq R_g^{hr}, \forall g, t, s \qquad (9)$$

$$P_{kts} = \frac{\theta_{kts}}{x_k}, \forall k, t, s \qquad (10)$$

$$-P_k^{max} \leq P_{kts} \leq P_k^{max}, \forall k, t, s \qquad (11)$$

$$0 \leq P^C_{rts} \leq P_{rts} \qquad (12)$$

$$0 \leq P^C_{wts} \leq P_{wts} \qquad (13)$$

$$\sum_{g \in G(n)} P_{gts} + \sum_{k \in K(n-)} P_{kts} - \sum_{k \in K(n+)} P_{kts}$$
$$= d_{nt} - \sum_{r \in R(n)} (P_{rts} - P^C_{rts})$$
$$- \sum_{w \in W(n)} (P_{wts} - P^C_{wts})$$
$$, \forall n, t, s \qquad (14)$$

*C. BESS and BESS-based VTL Formulation*

This sub-section presents the formulations for BESS and BESS-based VTL. Constraint (15) defines the BESS status. The charging and discharging power rate of BESS must be within the maximum physical limits as enforced in (16)-(17). Constraint (18) sets the boundaries of the BESS energy level. In (19), the BESS energy level calculation considers charging and discharging efficiencies. The nodal power balance equation (14) needs to be replaced by the updated constraint (20) to account for charging and discharging activities from the BESS.

$$u^c_{ets} + u^d_{ets} \leq 1, \forall e, t, s \qquad (15)$$

$$0 \leq P^c_{ets} \leq P_e^{c,max} u^c_{ets}, \forall e, t, s \qquad (16)$$

$$0 \leq P^d_{ets} \leq P_e^{d,max} u^d_{ets}, \forall e, t, s \qquad (17)$$

$$E_e^{min} \leq E_{ets} \leq E_e^{max}, \forall e, t, s \qquad (18)$$

$$E_{ets} = E_{e,t-1,s} + (\eta_e^c P^c_{ets} - P^d_{ets}/\eta_e^d) \Delta T \qquad (19)$$

$$\sum_{g \in G(n)} P_{gts} + \sum_{k \in K(n-)} P_{kts} - \sum_{k \in K(n+)} P_{kts}$$
$$= d_{nt} - \sum_{r \in R(n)} (P_{rts} - P^C_{rts})$$
$$- \sum_{w \in W(n)} (P_{wts} - P^C_{wts})$$
$$+ \sum_{e \in ES(n)} (P^c_{ets} - P^d_{ets}), \forall n, t, s \qquad (20)$$

Equations (21)-(22) are constraints that apply the VTL strategy during the optimization process. $ES(vt)$ is a set of two BESSs that are located at the two ends of a congested physical line, respectively, to ensure VT behavior for each $vt$.

$$\sum_{e \in ES(vt)} u^c_{ets} \leq 1, \forall vt, t, s \qquad (21)$$

$$\sum_{e \in ES(vt)} u^d_{ets} \leq 1, \forall vt, t, s \qquad (22)$$

TABLE I. MODEL DESCRIPTIONS AND FORMULATION SUMMARY

| Models | Descriptions | Equations |
|---|---|---|
| Base | The SSCUC model without any congestion mitigation scheme is a benchmark. | (1)-(14) |
| PT | A new physical line is added to the system and the SSCUC model. | (1)-(14)* |
| BESS | BESSs are added to the system and the SSCUC model. | (1)-(13), (15)-(20) |
| VTL | VT operation constraints are added to the SSCUC-BESS model. | (1)-(13), (15)-(22) |

* While SSCUC-PT and SSCUC-Base share the same formulation, the network for SSCUC-PT has additional physical line compared to SSCUC-Base.

Four different SSCUC optimization models for day-ahead generation scheduling can be formulated and implemented to evaluate the performance of various congestion mitigation schemes. They are explained in Table I.

## D. Market Analysis Metrics

Most US grid operators have adopted the locational marginal price (LMP) for market clearance. LMP is the marginal cost of supplying one additional MW of power to a given location. In the stochastic SCUC problem, it is equal to the weighted dual variable of the nodal power balance constraint. The LMP value can be used to calculate the load payment as in Equation (23), which is an important social welfare indicator.

$$Load\ Payment = \sum_t \sum_n \sum_s d_{nt} LMP_{nts}, \forall n, t, s \quad (23)$$

## III. CASE STUDIES

All four SSCUC problems, as defined in Table I, were solved using the Gurobi solver [13], which was coded with the Python-based Pyomo package [14]. An Anaconda Spyder environment with Python version 3.8.6 is used on an Intel(R) Core(TM) i5–3570 K CPU @ 3.40 GHz computer to develop the optimization program.

### A. Test System Description

We use the enhanced IEEE 24-bus test system, as shown in Figure 1, to compare the four SSCUC models and thus demonstrate the performance of the proposed SSCUC-VTL model. The enhanced test system is developed based on the 24-bus network described in [15]. All conventional coal-type generators on buses 2, 15, 16, and 23 are replaced by solar or wind generators. In the PT line case, a new line is added to connect buses 11 and 14, with identical line parameters as the existing line. In BESS-related cases, identical BESSs are added to bus 11 and bus 14.

### B. Uncertainty Modeling

A set of possible scenarios with the associated probabilities is provided as input to the SSCUC models. The scenarios are generated randomly by adding Gaussian noise as the forecast error to solar and wind generation data.

Three sets of error variation are generated for solar and wind data. The mean of the noise distribution curve is zero for both solar and wind. The standard deviation for solar is set based on the time of day [16] [17]. The standard deviation is set to 0 between 8 pm and 5 am, 0.05 between 9 am and 4 pm, and 0.02 for the rest of the day. The standard deviation for wind is set as 0.1 [18]. Fig. 2 shows the solar and wind generation profiles after adding the error variation to the base solar and wind generation. The base solar generation accounts for 40% of the peak load, while the base of the wind generation accounts for 30% of the peak load.

In this study, uncertainty is only introduced to solar and wind generation. The load profile remains consistent across all scenarios, mirroring the summer weekday load.

### C. SSCUC Model Simulation Results

The SSCUC model incorporates RES uncertainty into consideration by using scenarios representing selected RES levels with associated probability. This study considers three scenarios, with the possibility of scenario 1 as 25%, scenario 2 as 35%, and scenario 3 as 40%. Since the SSCUC model optimizes the unit status to work with all possible scenarios, the solution from the SSCUC model can differ from the solutions that are individually optimized by the Deterministic SCUC (DSCUC) model for a scenario in the scenario set.

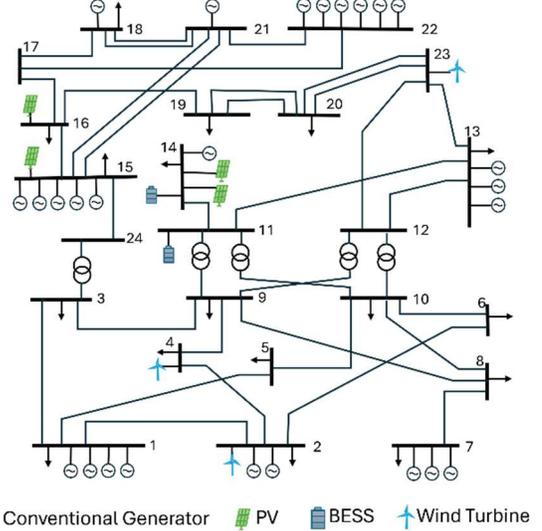

Figure 1. Topology of the enhanced IEEE 24-bus system.

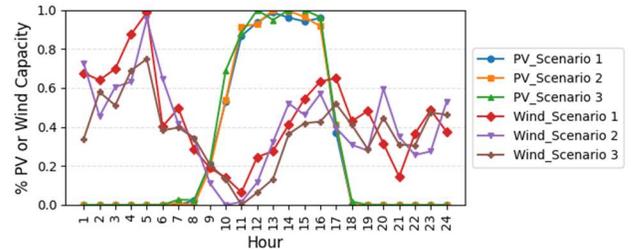

Figure 2. Illustration of the uncertainty of PV and wind generation.

Fig. 3 demonstrates the difference between SSCUC and DSCUC. Fig. 3(a) shows the unit commitment status for each generator over 24 hours, as solved by the SSCUC-VTL model. The SSCUC-VTL results provide an optimized solution across all scenarios, accounting for RES uncertainty. Fig. 3(b) shows the unit status and generation output obtained by the DSCUC model for scenario 1 in the SSCUC-VTL scenario set. Fig. 3(c) shows the unit status and generation output obtained by the DSCUC model for scenario 2 in the SSCUC-VTL scenario set. Fig. 3(d) shows the unit status and generation output obtained by the DSCUC model for scenario 3 in the SSCUC-VTL scenario set. Fig. 3(b)-(d) demonstrates that the DSCUC solution varies with changes in the renewable profiles. SSCUC-VTL provides a robust solution for all scenarios to resolve conflicts seen between DSCUC solutions.

Although the unit commitment status from SSCUC-VTL remains the same across all scenarios, network operating conditions, such as generation output and line loading, vary by scenario due to uncertainty in RES generation. Fig. 4 shows the different network loading levels for each scenario optimized by the SSCUC-VTL model. Different profiles are also present in the generator delivery level, the RES curtailment level, the BESS charging/discharging level, and the bus LMP.

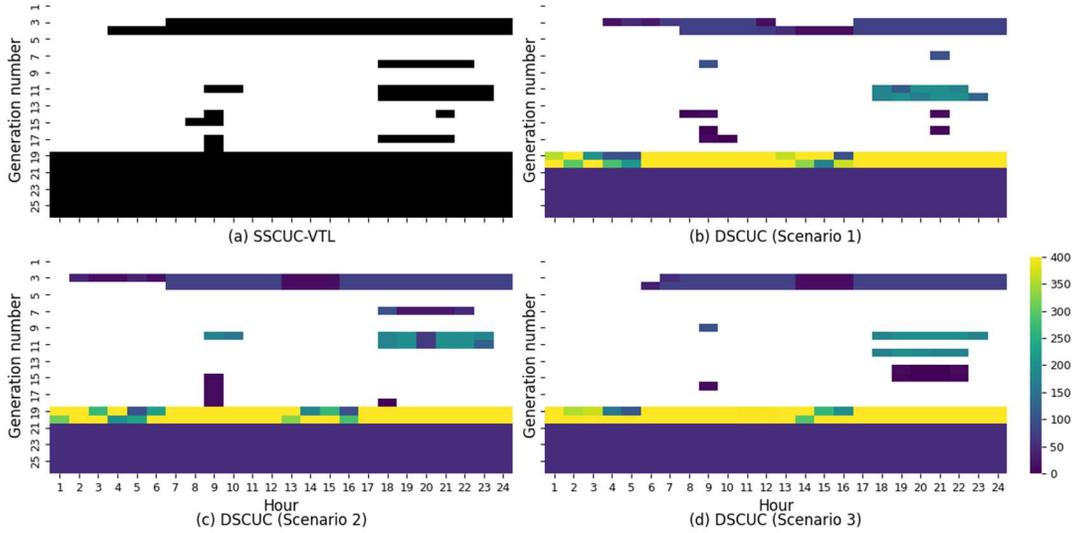

Figure 3. Illustration of generator unit commitment status and power outputs.

*D. SSCUC VTL Performance Analysis*

This section compares the SSCUC performance and benefits of VTL against the base case, physical line, and standalone BESS.

Table II compares the operational cost and load payment results of PT, standalone BESS, and VTL as a percentage of the base case results. Table III lists the total number of congested lines in the 24-bus system over the optimized 24-hour period for base, PT, standalone BESS, and VTL cases.

It is noticeable that the VTL scheme offers a significant operational cost reduction, which is comparable to the standalone BESS. One should also carefully note that VTL results in a higher load payment than all other cases. High LMP causes high load payments during certain hours. Although it cannot beat PT in terms of congestion reduction, the VTL strategy can alleviate congestion across all scenarios, whereas the BESS worsens congestion in all scenarios. More details can be found in Fig. 5, which displays the branch loading level of scenario 1 for base, PT, standalone BESS, and VTL cases.

Table IV compares the RES curtailment of PT, standalone BESS, and VTL as a percentage of the base case results for all scenarios. PT does not help with curtailment reduction. Both standalone BESS and VTL help reduce RES curtailment. There is no RES curtailment in scenario 3.

TABLE II. OPERATIONAL COST AND LOAD PAYMENT FOR VARIOUS CONGESTION MITIGATION SCHEMES

|  | Base | PT | BESS | VTL |
|---|---|---|---|---|
| **Operational cost** | 100% | 91% | 61% | 68% |
| **Load payment** | 100% | 91% | 87% | 105% |

TABLE III. SYSTEM CONGESTION IN EACH SCENARIO FOR VARIOUS CONGESTION MITIGATION SCHEMES

|  | Base | PT | BESS | VTL |
|---|---|---|---|---|
| **Scenario 1** | 8 | 4 | 10 | 8 |
| **Scenario 2** | 7 | 2 | 9 | 6 |
| **Scenario 3** | 6 | 0 | 9 | 5 |

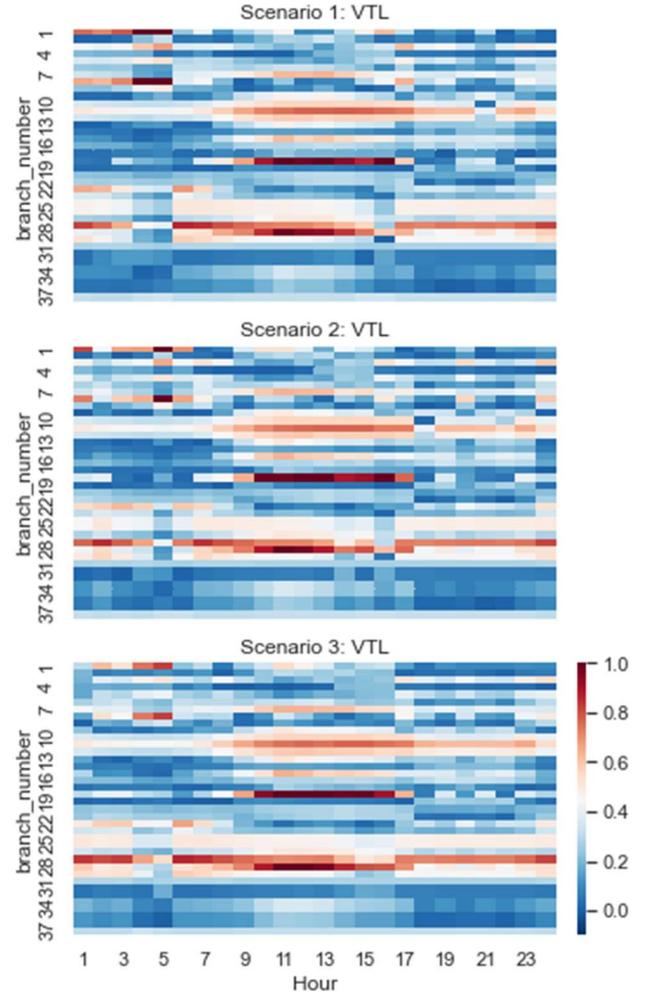

Figure 4. Branch loading level in each scenario of SSCUC-VTL.

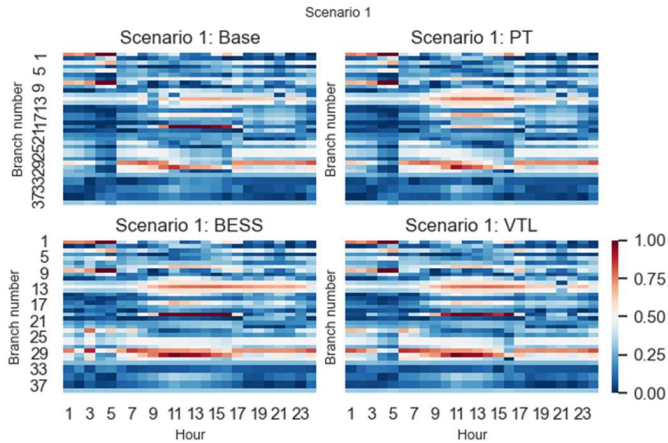

Figure 5. Branch loading level in each case for scenario 1.

TABLE IV. COMPARISON OF RES CURTAILMENT IN EACH SCENARIO FOR VARIOUS CONGESTION MITIGATION SCHEMES WITH SOLAR AND WIND PENETRATION

|  | Base | PT | BESS | VTL |
|---|---|---|---|---|
| **Scenario 1** | 100% | 100% | 92% | 95% |
| **Scenario 2** | 100% | 100% | 93% | 96% |
| **Scenario 3** | 0 | 0 | 0 | 0 |

## IV. CONCLUSION

This paper proposes a stochastic security-constrained unit commitment model for virtual transmission line applications, capturing the uncertainty of PV and wind generation through consideration of multiple potential scenarios. The SSCUC model is also suitable for incorporating standalone BESS and new physical transmission lines. An enhanced IEEE 24-bus test system is used to demonstrate the usability of the proposed SSCUC-VTL model. In the case study, the performance of the VTL is compared with the base case, the PT case, and the stand-alone BESS case. The introduction of RES uncertainty results in different power flows and LMP solutions for various scenarios. The results show that VTL can help reduce operation costs and RES curtailment significantly compared to the base and PT cases. VTL can provide more congestion relief than the BESS case. VTL can cause higher load payments than the Base, PT, and BESS cases.

Although this work theoretically proves the benefit of the SSCUC-VTL application, there are potential challenges in VTL's implementation regarding ownership regulation, market design, communication system design, and energy management control design, which require further studies.